\RequirePackage{fix-cm}
\documentclass[smallextended]{svjour3}       
\smartqed  
\usepackage{graphicx}
\begin{document}

\title{Protecting entanglement from correlated amplitude damping channel using weak measurement and quantum measurement reversal}


\author{Xing Xiao        \and
        Yao Yao \and
         Ying-Mao Xie\and
        Xing-Hua Wang \and
       Yan-Ling Li 
}


\institute{Xing Xiao \and Ying-Mao Xie \and Xing-Hua Wang \at
              College of Physics and Electronic Information, Gannan Normal University, Ganzhou 341000, China\\
                \email{xieyingmao@126.com}
           \and
           Yao Yao \at
           Microsystems and Terahertz Research Center,
China Academy of Engineering Physics,
Chengdu, Sichuan 610200, China
           \and
           Yan-Ling Li \at
              School of Information Engineering, Jiangxi University of Science and Technology, Ganzhou 341000, China  \\
                        \email{liyanling0423@gmail.com}
}

\date{Received: date / Accepted: date}

\maketitle

\begin{abstract}
Based on the quantum technique of weak measurement,
we propose a scheme to protect the entanglement from correlated amplitude damping decoherence. In contrast to the results of memoryless amplitude damping channel, we show that the memory effects play a significant role in the suppression of entanglement sudden death and protection of entanglement under severe decoherence. Moreover, we find that the initial entanglement could be drastically amplified by the combination of weak measurement and quantum measurement reversal even under the correlated amplitude damping channel. The underlying mechanism can be attributed to
the probabilistic nature of weak measurements.

\end{abstract}

\section{Introduction}
\label{intro} Quantum computation and quantum communication may well represent two of the most important breakthroughs in information technology since they can greatly enhance the computing speed and ensure the security of communication \cite{nielsen00}. These superiorities are based on the power of quantum entanglement, which is recognized as
an essential resource for quantum computation and quantum information. However,
entanglement is fragile and easily broken by environmental noise \cite{breuer02}. This would be the most limiting factor for the applications of entanglement in quantum computation \cite{divincenzo95}, quantum communication \cite{bennett93,gisin07}, quantum metrology \cite{giova06,giova11} and other quantum information processes. In this context, it is an extremely fundamental task to protect entanglement from the noise of channel.

An arbitrary physical process could be viewed, from the perspective of information theory, as a quantum channel which may be decoherent or not \cite{nielsen00,bennett98}. Usually, a quantum channel is defined mathematically as a completely positive, trace-preserving (CPTP) linear map on
density operators. The simplest example for quantum channel is the amplitude damping (AD) channel, which is a prototype model of a dissipative interaction between a qubit and its zero-temperature environment. If the AD noise acts identically and independently on each of the qubits that passes through the AD channel, we call it memoryless AD channel and the corresponding map is expressed as a tensor product of independent and identical CPTP maps. However, in many realistic scenarios, the Kraus operators of the AD channel map can not be expressed as a tensor product form, which means that the AD channel is memory or correlated among
consecutive uses \cite{holevo12,caruso14}. Note that quantum memory channel recently has attracted considerable attentions in information field since memory effects become unavoidable when increasing the transmission rate in quantum channels \cite{macch02,arrigo07,plenio07,arrigo13}.

Another interesting quantum technique which is emerging in recent years
is weak measurement (WM) \cite{koro06,katz08,kim09}. In contrast to 
the traditional von Neumann measurement, WM is more gentle in extracting the information from
the system and moreover, may enable the measured system to be alive (i.e., without completely collapsing towards an eigenstate). Thereby, a suitable quantum measurement reversal (QMR) could revive the state with a certain probability. Recently, many researches have indeed demonstrated that, under AD channel,  WM and QMR can enhance the fidelity of a single qubit, reverse entanglement change and even circumvent entanglement sudden death of two qubits and qutrits \cite{koro10,sun10,para11a,para11b,kim12,liyanling13,man12a,xiao13,man12b,wang14}. While most of the work produced so far has been restricted to the assumption that the AD channel acts identically and independently on the qubits (memoryless configuration in jargon). Memory or correlated amplitude damping (CAD) channels appear to be more reasonable and significant in quantum information theory. A naturally arising question is what would happen if the WM and QMR are applied to the case of CAD channel? 

Motivated by the above consideration, this study is to
discuss the role of WM and QMR in the protection of entanglement under two-qubit CAD channel, where the memory effects are characterized by a memory parameter $\eta$ which ranges from 0 to 1. The behaviors of entanglement under CAD channel are investigated with or without the assistance of WM and QMR. We show that WM and QMR enable the entanglement away from entanglement sudden death (ESD) \cite{yuting04,yuting09,almeida07} by choosing proper measurement strength of WM. Furthermore, we find that the initial entanglement could be drastically amplified by WM and QMR.
Particularly, for the memoryless and fully memory cases, the maximal achievable entanglement approaches to 1. Our results provide an active way to suppress decoherence and enhance the entanglement transmission under CAD channel, which is rather significant in quantum communications.

%

\begin{figure*}
  \includegraphics[width=1\textwidth]{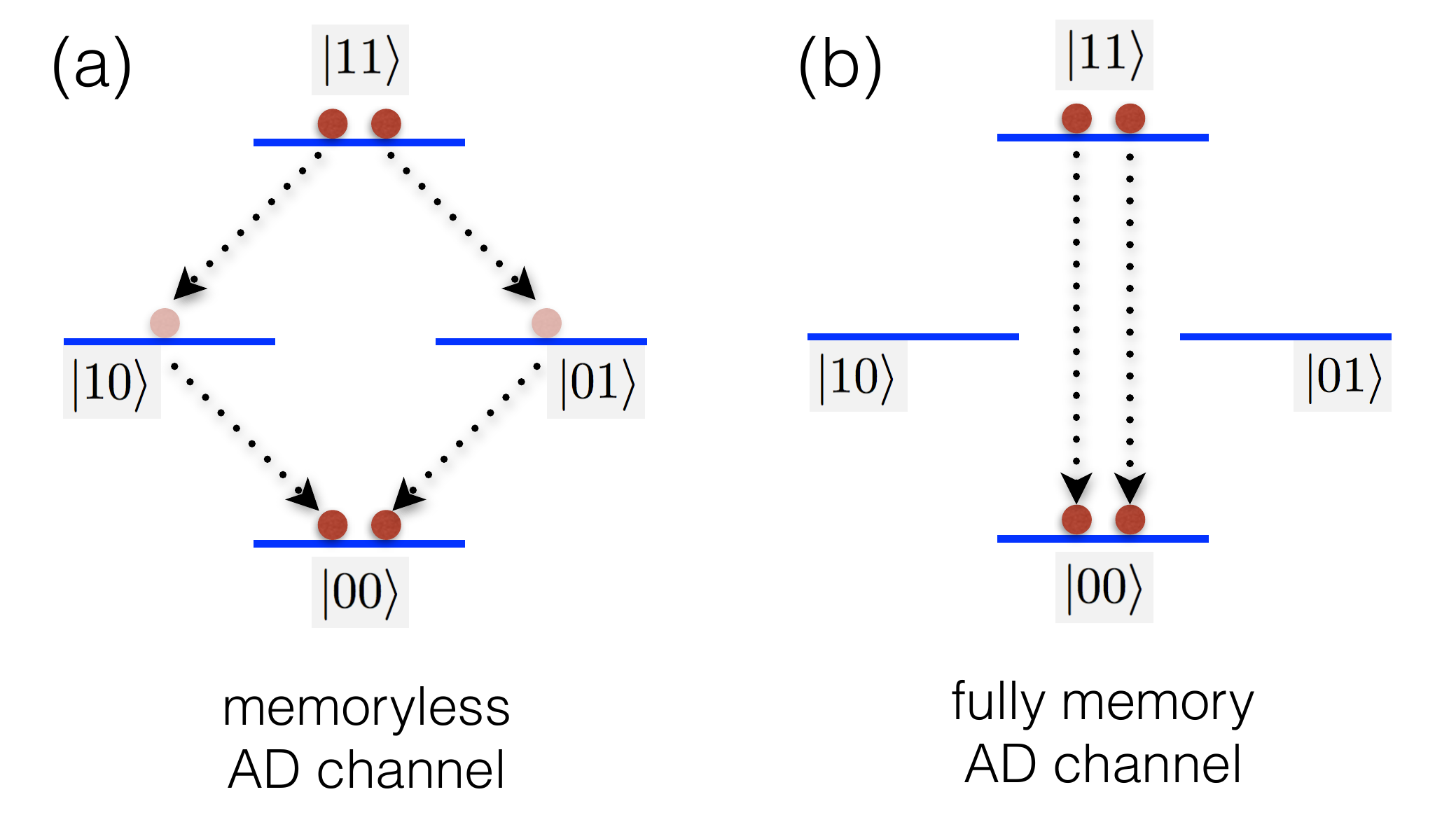}
\caption{(color online) Schematic illustrations of the relaxation mechanism. (a). In the memoryless AD channel, the two qubits decay independently. (b) In the fully memory AD channel, when a qubit
undergoes a relaxation process, the other qubit does the same.}
\label{Fig1}       
\end{figure*}

\section{Entanglement under CAD Channel}
\label{sec:1}

In the operator-sum representation, an arbitrary CPTP map $\mathcal{E}$ could be expressed as
$\mathcal{E}(\rho)=\sum_{i}E_{i}\rho E_{i}^{\dagger}$,
where $E_{i}$ are known as operation elements (also named as Kraus operators) for the quantum operation $\mathcal{E}$. For the AD channel, the corresponding Kraus operators are given as
\begin{eqnarray}
E_{0}=\left(\begin{array}{cc}1 & 0 \\0 & \sqrt{1-\gamma}\end{array}\right),\ E_{1}=\left(\begin{array}{cc}0 & \sqrt{\gamma} \\0 & 0\end{array}\right).
\label{e1}
\end{eqnarray}
Here we are using the orthonormal basis $|0\rangle$ and $|1\rangle$.
The parameter $\gamma\in[0, 1]$ is the probability of losing the system 
excitation into the environment, which is also known as the decoherence strength of the AD channel.
Note that for the two-qubit memoryless AD channel, the total evolution map is expressed as a tensor product of the above CPTP map: $\mathcal{E}^{(2)}(\rho)=\mathcal{E}^{\otimes2}(\rho)$.
However, when the tensorial decomposition isn't applicable, then the channel is usually regarded as memory or correlated channel. As proved in Ref. \cite{yeo03}, a CAD channel could be written as
\begin{eqnarray}
\label{e2}
\mathcal{E}_{\rm CAD}(\rho)&=&(1-\eta)\mathcal{E}_{\rm AD}^{\otimes2}(\rho)+\eta\mathcal{E}_{\rm FCAD}(\rho),\\
&=&(1-\eta)\sum_{i,j=0}^{1}E_{ij}\rho E_{ij}^{\dagger}+\eta\sum_{k=0}^{1}A_{k}\rho A_{k}^{\dagger},\nonumber
\end{eqnarray}
where the subscripts ``AD'', ``CAD'' and ``FCAD'' denote ``memoryless AD'', ``correlated AD'' and ``fully correlated AD'', respectively.
Obviously, $\eta\in[0,1]$ is the memory parameter. We can recover the memoryless AD channel by setting $\eta=0$ and
obtain the FCAD channel $\mathcal{E}_{\rm FCAD}$ if $\eta=1$. The explicit expression of the Kraus operators
$A_{k}$ are determined by solving the correlated Lindblad equation \cite{arshed13}, which gives the following formalism 
\begin{eqnarray}
A_{0}=\left(\begin{array}{cccc}1 & 0 & 0 & 0 \\0 & 1 & 0 & 0 \\0 & 0 & 1 & 0 \\0 & 0 & 0 & \sqrt{1-\gamma}\end{array}\right), A_{1}=\left(\begin{array}{cccc}0 & 0 & 0 & \sqrt{\gamma} \\0 & 0 & 0 & 0 \\0 & 0 & 0 & 0 \\0 & 0 & 0 & 0\end{array}\right).
\label{e3}
\end{eqnarray}
In contrast to the memoryless AD channel,  where the damping of two qubits occurs independently from $|1\rangle$ to $|0\rangle$, as shown in Fig. \ref{Fig1}a, the FCAD only allows the synchronous transition between $|1\rangle$ and $|0\rangle$, i.e., the relaxation from $|11\rangle$ to $|00\rangle$ as depicted in Fig. \ref{Fig1}b.

In this paper, we focus on the CAD channel of Eq. (\ref{e2}). First we consider the behavior of entanglement under CAD channel and then discuss the influence of WM and QMR on the entanglement protection. Let us suppose that the initial state is prepared in a maximally entangled state
\begin{equation}
|\psi\rangle=\alpha|00\rangle+\beta|11\rangle.
\label{e4}
\end{equation}
with $|\alpha|^2+|\beta|^2=1$.
After it goes through the CAD channel, the initial pure state inevitably evolves into a mixed state:
$\rho_{\rm CAD}=\mathcal{E}_{\rm CAD}(|\psi\rangle\langle\psi|)$. In the standard product basis $\{|ij\rangle,i,j=0,1\}$, the non-zero elements of $\rho_{\rm CAD}$ are
\begin{eqnarray}
\label{e5}
\rho_{\rm CAD}^{11}&=&|\alpha|^2+(\overline{\eta}\gamma^2+\eta\gamma)|\beta|^2,\\
\rho_{\rm CAD}^{22}&=&\rho_{\rm CAD}^{33}=\overline{\eta}\gamma\overline{\gamma}|\beta|^2,\nonumber\\
\rho_{\rm CAD}^{44}&=&(\overline{\eta}\ \overline{\gamma}^2+\eta\overline{\gamma})|\beta|^2,\nonumber\\
\rho_{\rm CAD}^{14}&=&\rho_{\rm CAD}^{41^{*}}=(\overline{\eta}\ \overline{\gamma}+\eta\sqrt{\overline{\gamma}})\alpha\beta^{*},\nonumber
\end{eqnarray}
where we have used the notation $\overline{o}\equiv1-o$ with $o=\gamma, \eta, p, q$.

\begin{figure*}
  \includegraphics[width=0.9\textwidth]{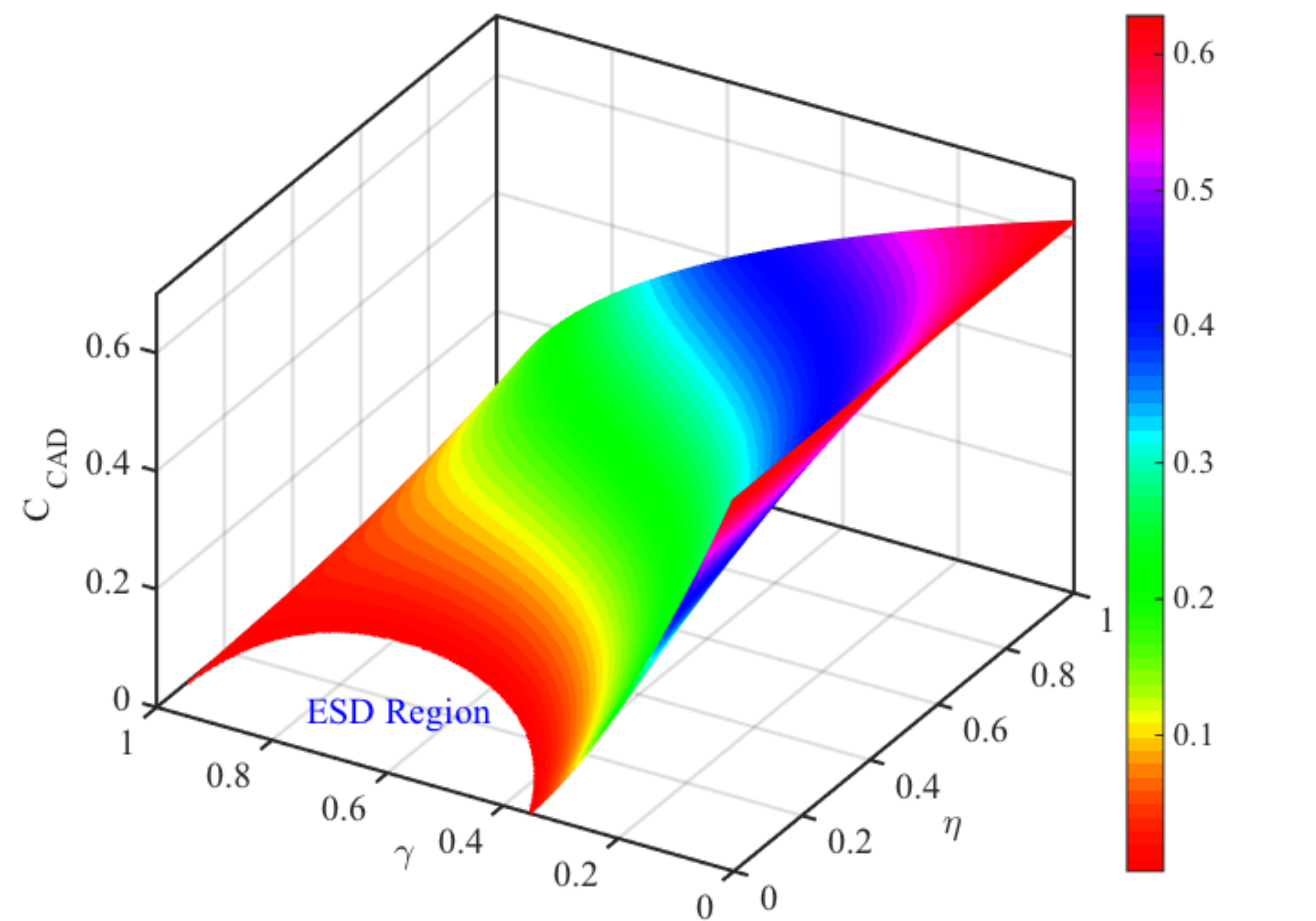}
\caption{(color online) Concurrence $C_{\rm CAD}$ as
a function of $\gamma$ and $\eta$ with $\alpha=1/3$. The semicircle in $\gamma-\eta$ plane indicates the region of ESD.}
\label{Fig2}       
\end{figure*}


In order to follow the two-qubit entanglement under CAD channel, we
adopt concurrence to measure the entanglement \cite{wootters98}. It is
defined as
$C_{\rho}=\max\{0,\sqrt{\lambda_{1}}-\sqrt{\lambda_{2}}-\sqrt{\lambda_{3}}-\sqrt{\lambda_{4}}\}$,
where $\lambda_{i},(i=1,2,3,4)$ are the eigenvalues, in decreasing
order, of matrix $\tilde{\rho}=\rho(\sigma_{y}\otimes\sigma_{y})\rho^*(\sigma_{y}\otimes\sigma_{y})$; $\sigma_{y}$ is the Pauli spin matrix and the asterisk denotes
complex conjugation. In
particular, the concurrence, for the density matrix of Eq. (\ref{e5}), is
given by
\begin{equation}
\label{e6}
C_{\rm CAD}=2\max\left\{0, \Delta_{\rm CAD}\equiv(\overline{\eta}\ \overline{\gamma}+\eta\sqrt{\overline{\gamma}})|\alpha\beta|-\overline{\eta}\gamma\overline{\gamma}|\beta|^2\right\}.
\end{equation}

The behaviors of concurrence as a function of dimensionless parameters $\gamma$ and $\eta$ are illustrated in Fig. \ref{Fig2} with $\alpha=1/3$. It is noted that for the memoryless AD channel ($\eta=0$), the concurrence reduces to $C_{\rm AD}=2\max\{0,\overline{\gamma}|\beta|(|\alpha|-\gamma|\beta|)\}$, which degrades monotonically with the increasing decoherence strength $\gamma$. Furthermore, it is essential to point out that $C_{\rm AD}$ would experience ESD. While considering the memory effects in CAD channel, the results are more interesting. 
It seems that the memory effects have a two-fold effect on the entanglement: the one is suppressing the entanglement degradation and the other is inducing the entanglement revival after a short period of ESD. For example, when $\alpha=1/3$ and $\eta=0.2$, the entanglement decays and suffers ESD in the region $0.465<\gamma<0.932$, while a small part of entanglement is finally revived for $0.932<\gamma<1$. This result is similar to the entanglement dynamics in non-Markovian environments where memory effects also induce the entanglement revival \cite{bellomo07}. When the memory effects are strong enough, i.e., $\eta\geq\eta_{\rm c}$, the phenomena of ESD is completely eliminated, where the critical value $\eta_{\rm c}$ is given by
\begin{equation}
\eta_{\rm c}=\frac{\gamma\sqrt{\overline{\gamma}}-|\frac{\alpha}{\beta}|\sqrt{\overline{\gamma}}}{|\frac{\alpha}{\beta}|(1-\sqrt{\overline{\gamma}})+\gamma\sqrt{\overline{\gamma}}}.
\end{equation}

\begin{figure*}
  \includegraphics[width=1\textwidth]{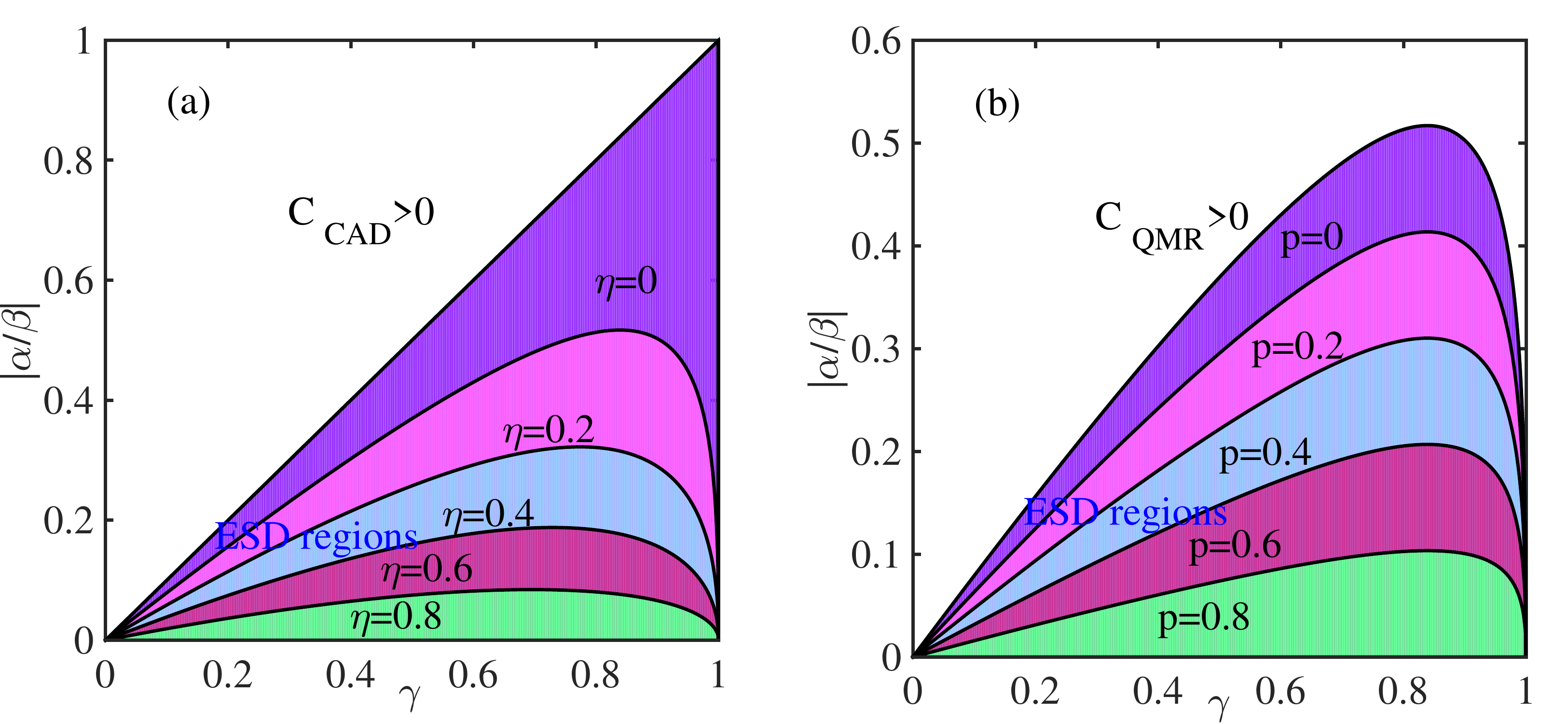}
\caption{(color online) (a) Diagram of the regions of ESD for CAD channel under different memory parameters without WM and QMR. (b) Diagram of the regions of ESD for CAD channel with the help of WM and QMR for different measurement strength of WM and given $\eta=0.2$. In both diagrams, white regions denote non-zero entanglement while shadow regions indicate zero entanglement.}
\label{Fig3}       
\end{figure*}

It is well known that ESD has significant impact on the quest to build large entangled states for applications in quantum information science and fundamental quantum physics. The condition of ESD under CAD channel needs to be clarified further.
It is clear from Eq. (\ref{e6}) that the condition of ESD is $\Delta_{\rm CAD}=0$, which means
\begin{equation}
|\frac{\alpha}{\beta}|<\frac{\overline{\eta}\gamma\sqrt{\overline{\gamma}}}{\overline{\eta}\sqrt{\overline{\gamma}}+\eta}.
\label{e7}
\end{equation}
It is interesting to note that the stronger the memory effects, $\eta\rightarrow1$, the smaller area of the ESD region, as shown in Fig. \ref{Fig3}a. When the CAD channels are fully correlated $\eta=1$, the phenomena of ESD disappears for arbitrary initial parameters $\alpha$ and $\beta$.

\section{Protecting Entanglement by WM and QMR}
\label{sec:2}

Then we turn to discuss the effects of WM and QMR on the entanglement transmission.
Before the qubits suffer the CAD noise, they are subject to a prior WM, which is described by a non-unitary quantum operation
\begin{equation}
\mathcal{M}_{\rm WM}=\left(\begin{array}{cc}1 & 0 \\0 & \sqrt{1-p}\end{array}\right)\otimes\left(\begin{array}{cc}1 & 0 \\0 & \sqrt{1-p}\end{array}\right),
\label{e8}
\end{equation}
where $p$ is the measurement strength of WM. Note that the WM doesn't completely collapse the state towards $|00\rangle$ or $|11\rangle$, which means that the measured state is still recoverable by proper operations, e.g., QMR.

After the CAD channel, a post QMR is performed on the qubits. The QMR is also a non-unitary operation which gives
\begin{equation}
\mathcal{M}_{\rm QMR}=\left(\begin{array}{cc}\sqrt{1-q} & 0 \\0 & 1\end{array}\right)\otimes\left(\begin{array}{cc}\sqrt{1-q} & 0 \\0 & 1\end{array}\right),
\label{e9}
\end{equation}
where $q$ is the measurement strength of QMR. The final state after the sequential WM, CAD channel and QMR is
\begin{equation}
\rho_{\rm QMR}=\mathcal{M}_{\rm QMR}\left[\mathcal{E}_{\rm CAD}\left(\mathcal{M}_{\rm WM}|\psi\rangle\langle\psi|\mathcal{M}_{\rm WM}^{\dagger}\right)\right]\mathcal{M}_{\rm QMR}^{\dagger},
\label{e10}
\end{equation}
which has the following non-zero elements:
\begin{eqnarray} 
\label{e11}
\rho_{\rm QMR}^{11}&=&\frac{\overline{q}^2U}{N(|\alpha|^2+\overline{p}^2|\beta|^2)},\nonumber\\
\rho_{\rm QMR}^{22}&=&\rho_{\rm QMR}^{33}=\frac{\overline{q}V}{N(|\alpha|^2+\overline{p}^2|\beta|^2)},\\
\rho_{\rm QMR}^{44}&=&\frac{W}{N(|\alpha|^2+\overline{p}^2|\beta|^2)},\nonumber\\
\rho_{\rm QMR}^{14}&=&\rho_{\rm QMR}^{41^{*}}=\frac{\overline{q}X}{N(|\alpha|^2+\overline{p}^2|\beta|^2)},\nonumber
\end{eqnarray}
with $U=|\alpha|^2+\overline{p}^2(\overline{\eta}\gamma^2+\eta\gamma)|\beta|^2$, $V=\overline{p}^2\overline{\eta}\gamma\overline{\gamma}|\beta|^2$, $W=\overline{p}^2(\overline{\eta}\ \overline{\gamma}^2+\eta\overline{\gamma})|\beta|^2$ and $X=\overline{p}(\overline{\eta}\ \overline{\gamma}+\eta\sqrt{\overline{\gamma}})\alpha\beta^{*}$. $N=(\overline{q}^2U+2\overline{q}V+W)/(|\alpha|^2+\overline{p}^2|\beta|^2)$ is the normalization factor.

The entanglement of $\rho_{\rm QMR}$ is calculated to be
\begin{equation}
\label{e12}
C_{\rm QMR}=\max\left\{0,\Delta_{\rm QMR}\equiv\frac{2\overline{q}(|X|-V)}{\overline{q}^2U+2\overline{q}V+W}\right\}.
\end{equation}
In order to make sure that $\Delta_{\rm QMR}$ reaches its maximal value, we need to optimize the strength $q$ of QMR. Using the fact that $a+b\geq2\sqrt{ab}$, ($a,b>0$), we have the following inequality
\begin{equation}
\label{e13}
\Delta_{\rm QMR}\leq\frac{|X|-V}{V+\sqrt{UW}},
\end{equation}
where the equality sign stands if and only if 
\begin{equation}
\label{e14}
\overline{q}=\sqrt{\frac{W}{U}}.
\end{equation}
Substituting the optimal strength of QMR, i.e.,Eq. (\ref{e14}), into the expression of $\Delta_{\rm QMR}$, we get
\begin{equation}
\label{e15}
\Delta_{\rm QMR}^{\rm opt}=\frac{(\overline{\eta}\ \overline{\gamma}+\eta\sqrt{\overline{\gamma}})|\alpha|-\overline{p}\ \overline{\eta}\gamma\overline{\gamma}|\beta|}{\overline{p}\ \overline{\eta}\gamma\overline{\gamma}|\beta|+\sqrt{[|\alpha|^2+\overline{p}^2(\overline{\eta}\gamma^2+\eta\gamma)|\beta|^2](\overline{\eta}\ \overline{\gamma}^2+\eta\overline{\gamma})}}.
\end{equation}

With the above equation in mind, we are now ready to answer the question asked in the introduction, i.e., what would happen if the WM and QMR are applied to the case of CAD channel?  Remarkably, we can draw two important conclusions from the result in
Eq. (\ref{e15}). First, the region of ESD is greatly reduced since the original ESD condition is modified by $1-p$\begin{equation}
|\frac{\alpha}{\beta}|<\frac{(1-p)\overline{\eta}\gamma\sqrt{\overline{\gamma}}}{\overline{\eta}\sqrt{\overline{\gamma}}+\eta}.
\end{equation}
This means that WM and QMR can be used to effectively circumvent ESD if $p\rightarrow1$. Figure \ref{Fig3}b shows the influence of WM and QMR on the ESD region of CAD channel with the memory strength $\eta=0.2$. As one might expected, the ESD region decreases with the increase of WM strength $p$.

\begin{figure*}
  \includegraphics[width=1\textwidth]{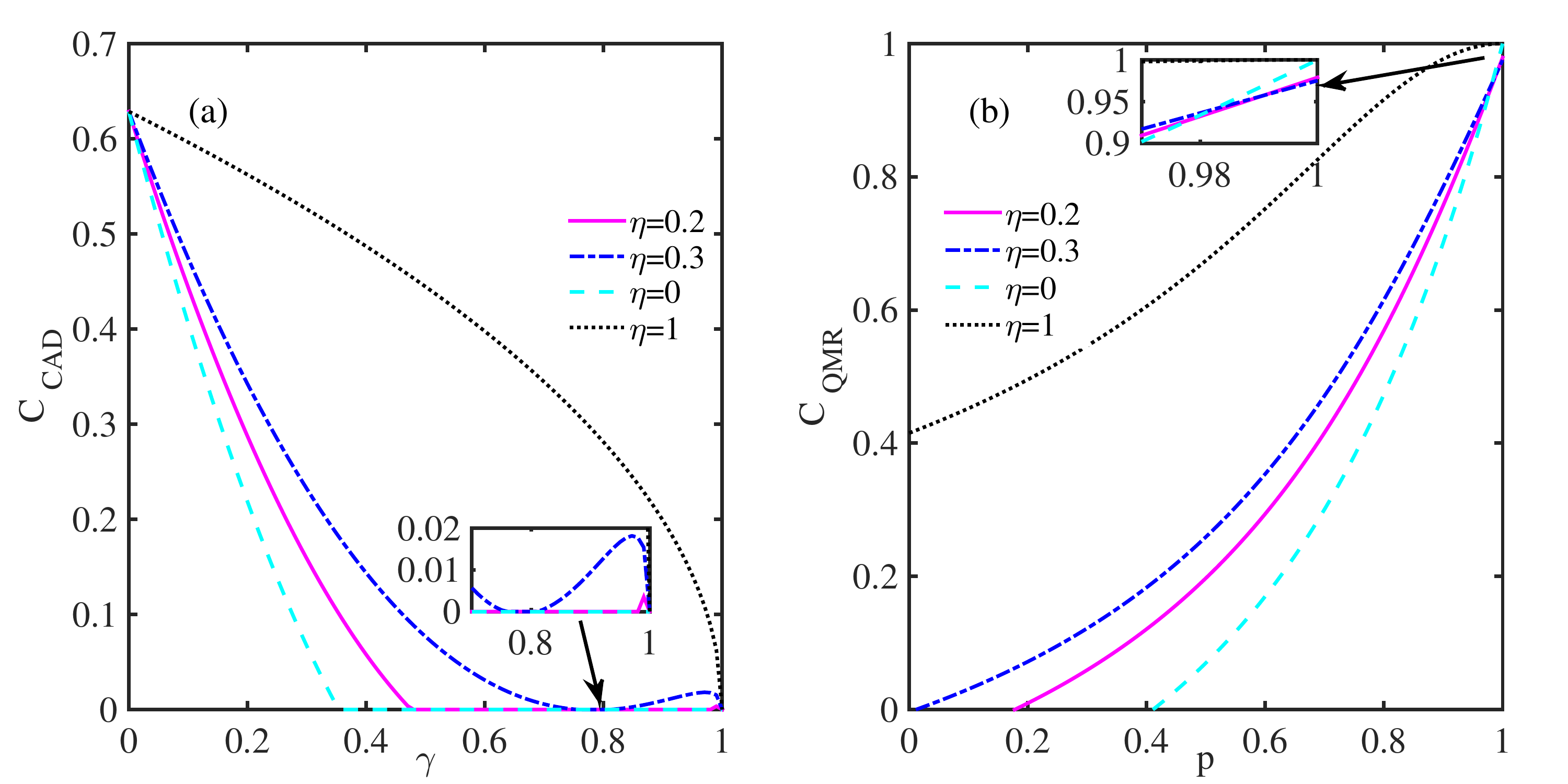}
\caption{(color online) (a) Concurrence as a function of decoherence strength $\gamma$ for different memory parameters. The insert shows that memory effects induce entanglement revival after a short period of ESD. (b) Concurrence as a function of measurement strength of WM for different memory parameters where we have assumed $\gamma=0.6$. The insert shows that the maximally achievable entanglement is 1 for $\eta=0,1$ and smaller than 1 for $\eta\neq0,1$.}
\label{Fig4}       
\end{figure*}

%
%

Second, the maximal value of $\Delta_{\rm QMR}=\frac{\overline{\eta}\sqrt{\overline{\gamma}}+\eta}{\sqrt{(\overline{\eta}\ \overline{\gamma}+\eta)}}$ is obtained when the strength of the WM $p=1$. It is surprising to note that the maximum concurrence is independent of the initial state parameters $\alpha$ and $\beta$. At the first glance, one might conjecture that entanglement has been created by WM and QMR, which are local operations. 
In fact, it is just an illusion because the success probability is zero when $\alpha=0$ or $\beta=0$, which means that one cannot create entanglement only by local operations if there is no entanglement initially. However, entanglement could be amplified by WM and QMR for initially entangled states. In Fig. \ref{Fig4}, we show the amplification of entanglement from CAD channel by using WM and QMR. In contrast to the results of pure CAD channel (see Fig. \ref{Fig4}a), the entanglement assisted by WM and QMR highly depends on the measurement strength of WM.
To demonstrate the ability of WM and QMR to amplify entanglement even under severe decoherence, we chose $\alpha=1/3$ and $\gamma=0.6$ as an example.  Note that under these parameters, the concurrence $C_{\rm CAD}$ would suffer ESD if the memory parameter is small. However, we show that WM and QMR can eliminate ESD for the WM strength $p$ larger than a critical value $p_{\rm c}=|\frac{\alpha}{\beta}|(1-\frac{\overline{\eta}\sqrt{\overline{\gamma}}+\eta}{\overline{\eta}\gamma\sqrt{\overline{\gamma}}})$. Moreover,it is remarkable to find that the entanglement could be dramatically amplified and may be even larger than initial entanglement with the combined action of WM and QMR.  For instance, the initial concurrence is 0.6285 when $\alpha=1/3$, while the achievable concurrence is obviously larger than it, as shown in Fig. \ref{Fig4}b. Nevertheless, since WM and QMR are non-unitary operations, the amplification is obtained at the expense of low success probability, i.e, the larger the amplified entanglement, the lower the success probability.  

\begin{figure*}
  \includegraphics[width=0.8\textwidth]{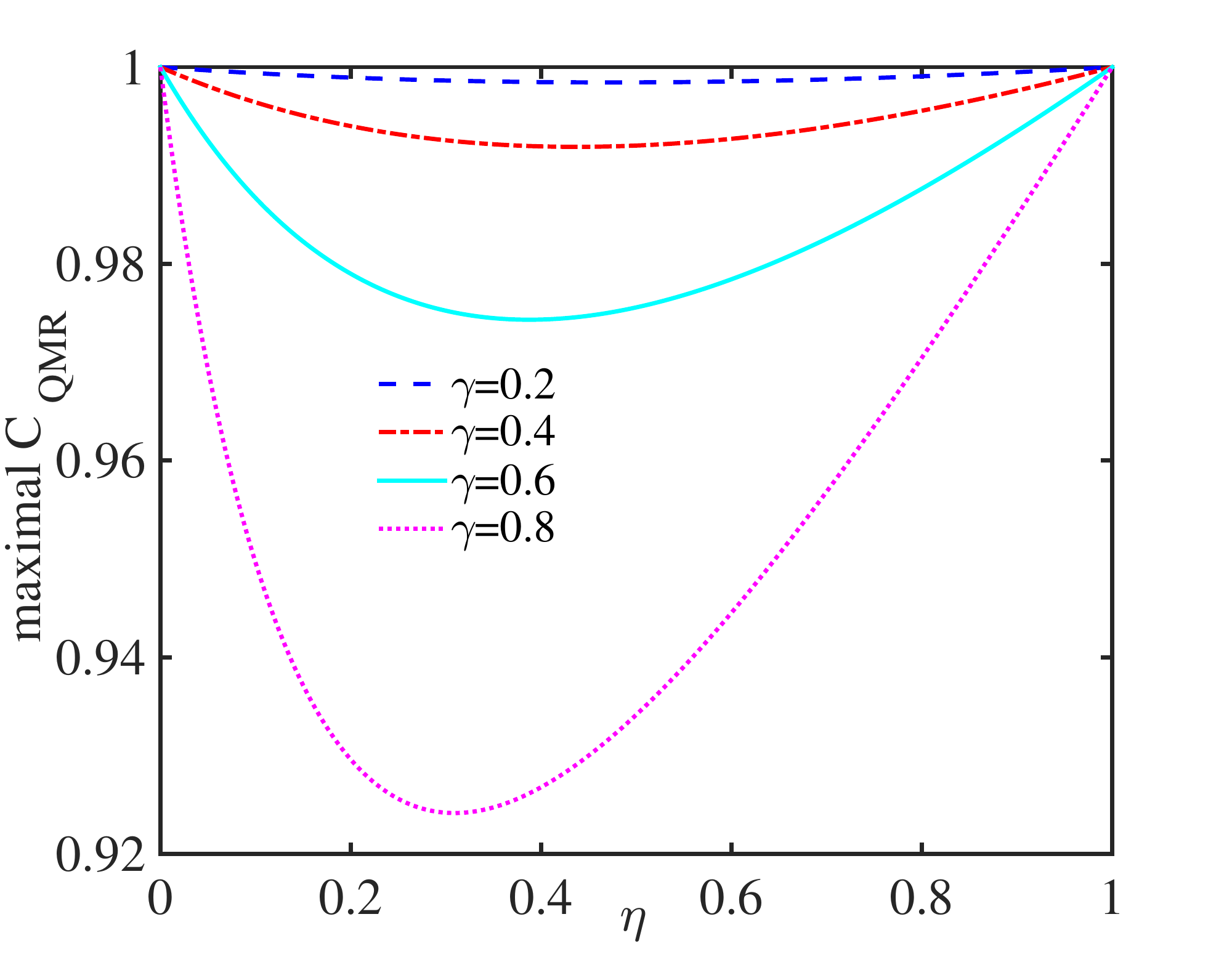}
\caption{(color online) Maximally achievable concurrence $C_{\rm QMR}$ as a function of memory parameter $\eta$ for different decoherence strengths.}
\label{Fig5}       
\end{figure*}

On the other hand, we would like to point out that two limiting cases are interesting. (i) When the CAD channel reduces to a memoryless AD channel ($\eta=0$), our results naturally reduce to those obtained in Ref. \cite{kim12} and the maximal value of $\Delta_{\rm QMR}$ is 1. (ii) For the fully correlated case ($\eta=1$),  the maximal value of $\Delta_{\rm QMR}$ is also 1. While for the intermediate cases $0<\eta<1$, the maximal value of $\Delta_{\rm QMR}$ is less than 1, as shown in the insert of Fig. \ref{Fig4}b. The underlying mechanism could be understood as follows:
from Eq. (\ref{e8}) we note that the key function of WM is projecting the quantum state to $|00\rangle$ with probability $p$ because $|00\rangle$ is immune to both memoryless AD and fully CAD decoherence. After the noise channel, the QMR is performed which is aimed to retrieve the entanglement as most as possible. For both memoryless AD and fully CAD channels, the QMR described by Eq. (\ref{e9}) is effective, but for the general CAD channel (i.e., $\eta\neq0,1$), the QMR of Eq. (\ref{e9}) is not optimal. The reason is that during the reversal procedure, $\mathcal{\rm QMR}$ can't exactly distinguish the memoryless transition $|11\rangle\rightarrow(|10\rangle, |01\rangle)\rightarrow|00\rangle$ and fully memory transition $|11\rangle\rightarrow|00\rangle$. Therefore, the entanglement couldn't be amplified to 1 by the operation of QMR if the mixing of two transitions exist, as shown in Fig. \ref{Fig5}. Naturally, one may wonder whether there is another type of QMR that enables us to distinguish these two transitions accurately. Unfortunately, so far we cannot answer this problem affirmatively, which is left as an open question. 

\section{Discussion and Conclusion}
\label{sec:3}

Before conclusion, we would like to discuss the experimental realizations of WM and QMR.
Generally speaking, WM is a particular type of POVM formalism. As shown in Ref. \cite{nielsen00}, any POVM could be realized with the combination of projective measurement and unitary dynamics of a composite system including ancillary system and target system. 
Hence, the performance of WM on the target qubit is equivalent to the action of von Neumann projective measurement on the ancilla qubit which is previously coupled to it. Alternatively, the WM also could be realized with a single operation, as shown in Refs. \cite{kim09,katz08}. This is important in realistic experiments since the technique of controlling two qubits and their coupling is difficult. On the other hand, the operation time is also a significant resource due to the limited coherence time. For example, as shown in Refs. \cite{kim09,kim12}, the WM can be implemented with a Brewster-angle glass plate (BAGP) for photon-polarization qubit because the BAGP probabilistically rejects vertical polarization ($|1\rangle$ state) and completely transmits horizontal polarization ($|0\rangle$ state), which exactly functions as the measurement depicted by Eq. (8). Meanwhile, the QMR could be decomposed as a sequential operations: bit-flip (a half-wave plate for polarization qubit), WM and bit-flip, which is also entirely feasible in experiment.


In summary, we have proposed a scheme to protect the entanglement from CAD decoherence utilizing WM and QMR. The two-fold effect of memory effects on the entanglement has been revealed. Remarkably, we further show that the combination of WM and QMR can effectively reduce the ESD region and even amplify the entanglement. Our work extends the ability of WM and QMR as a new technique in various quantum information processing tasks, particularly, when the research targets are subject to CAD noise. Moreover, our investigation also gives an expectation of developing optimal WM and QMR that can distinguish the memoryless transition and fully memory transition perfectly, which would be more practical for protecting entanglement in a realistic CAD channel.

\begin{acknowledgements}
This work is supported by the Funds of the National Natural Science Foundation of China under Grant Nos. 11247006 and 11365011.
\end{acknowledgements}



\begin{thebibliography}{}

\bibitem{nielsen00}Nielsen, M.A., Chuang, I.L.: Quantum Computation and Quantum Information. Cambridge: Cambrige University
Press (2000)
\bibitem{breuer02}Breuer, H.P., Petruccione, F.: The Theory of Open Quantum Systems. Oxford: Oxford University Press (2002)
\bibitem{divincenzo95}DiVincenzo, D.P.: Quantum computation. Science, 270, 255-261 (1995)


\bibitem{bennett93}Bennett, C.H. et al.: Teleporting an unknown quantum state via dual
classical and Einstein-Podolsky-Rosen channels. Phys. Rev. Lett.
\textbf{70}, 1895-1899 (1993)


\bibitem{gisin07}Gisin, N., Thew, R.: Quantum communication. Nat. Photonics. \textbf{1}, 165-171 (2007)


\bibitem{giova06}Giovannetti, V., Lloyd, S.,  Maccone, L.: Quantum metrology. Phys. Rev.
Lett. \textbf{96}, 010401 (2006)

\bibitem{giova11}Giovannetti, V., Lloyd, S.,  Maccone, L.: Advances in quantum metrology.
Nat. Photonics. \textbf{5}, 222-229 (2011)

\bibitem{bennett98}Bennett, C.G., Shor, P.W.: Quantum information theory. IEEE Trans. Inform. Theory \textbf{44}, 2724-2742 (1998)

\bibitem{holevo12}Holevo, A.S., Giovannetti, V.:  Quantum channels and their entropic characteristics. Rep. Prog. Phys. \textbf{75}, 046001 (2012)

\bibitem{caruso14}Caruso, F., Giovannetti, V., Lupo, C., Mancini, S.: Quantum channels and memory effects. Rev. Mod. Phys. \textbf{86}, 1203 (2014)

\bibitem{macch02}Macchiavello, C., Palma, G.M.: Entanglement-enhanced information transmission over a quantum channel with correlated noise. Phys. Rev. A \textbf{65}, 050301(R) (2002)

\bibitem{arrigo07}D'Arrigo, A., Benenti, G., Falci, G.: Quantum capacity of dephasing channels with memory. New J. Phys. \textbf{9}, 310 (2007)

\bibitem{plenio07}Plenio. M.B., Virmani, S.: Spin chains and channels with memory. Phys. Rev. Lett. \textbf{99}, 120504 (2007)

\bibitem{arrigo13}D'Arrigo, A., Benenti, G., Falci, G., Macchiavello, C.: Classical and quantum capacities of a fully correlated amplitude damping channel. Phys. Rev. A \textbf{88}, 042337 (2013)


\bibitem{koro06}Korotkov, A.N., Jordan, A.N.: Undoing a weak quantum measurement of a
solid-state qubit. Phys. Rev. Lett. \textbf{97}, 166805 (2006)

\bibitem{katz08}Katz, N. et al.: Reversal of the weak measurement of a quantum state in a
superconducting phase qubit. Phys. Rev. Lett. \textbf{101}, 200401
(2008)
\bibitem{kim09}Kim, Y.S., Cho, Y.W., Ra, Y.S., Kim, Y.H.: Reversing the weak quantum
measurement for a photonic qubit. Opt. Express \textbf{17},
11978-11985 (2009)

\bibitem{koro10}Korotkov, A.N., Keane, K.: Decoherence suppression by quantum measurement
reversal. Phys. Rev. A \textbf{81}, 040103(R) (2010)

\bibitem{sun10}Sun, Q.Q., Al-Amri, M., Davidovich, L., Zubairy, M.S.: Reversing entanglement change by a weak measurement. Phys. Rev. A \textbf{82}, 052323 (2010)

\bibitem{para11a}Paraoanu, G.S.: Partial measurements and the realization of quantum-mechanical counterfactuals. Found. Phys. \textbf{41}, 1214-1235 (2011).

\bibitem{para11b}Paraoanu, G.S.: Generalized partial measurements. EPL (Europhysics Letters) \textbf{93}, 64002 (2011).

\bibitem{kim12}Kim, Y.S., Lee, J.C., Kwon, O., Kim, Y.H.: Protecting entanglement from decoherence using weak measurement and quantum measurement reversal. Nat. Phys.
\textbf{8}, 117 (2012)

\bibitem{liyanling13}Li, Y.L., Xiao, X.: Recovering quantum correlations from amplitude damping decoherence by weak measurement reversal.
Quantum Information Processing \textbf{12}, 3067-3077 (2013).
\bibitem{man12a}Man, Z.X., Xia, Y.J., An, N.B.: Manipulating entanglement of two qubits in a common environment by means of weak measurements and quantum measurement reversals. Phys. Rev. A \textbf{86}, 012325 (2012)

\bibitem{xiao13}Xiao, X., Li, Y.L.: Protecting qutrit-qutrit entanglement by weak measurement and reversal. Eur. Phys. J. D \textbf{67}, 204 (2013)
\bibitem{man12b}Man, Z.X., Xia, Y.J., An, N.B.: Enhancing entanglement of two qubits undergoing independent decoherences by local pre- and postmeasurements. Phys. Rev. A \textbf{86}, 052322 (2012)


\bibitem{wang14}Wang, S.C., Yu, Z.W., Zou, W.J., Wang, X.B.: Protecting quantum states from decoherence of finite temperature using weak measurement. Phys. Rev. A \textbf{89}, 022318 (2014)

\bibitem{yuting04}Yu, T., Eberly, J.H.: Finite-time disentanglement via spontaneous emission. Phys. Rev. Lett. \textbf{93}, 140404
(2004)
\bibitem{yuting09}Yu, T., Eberly, J.H.: Sudden death of entanglement.
Science \textbf{323}, 598-601 (2009)
\bibitem{almeida07}Almeida, M.P. et al.: Environment-induced sudden death of entanglement.
Science \textbf{316}, 579-582 (2007)

\bibitem{yeo03}Yeo, Y., Skeen, A.: Time-correlated quantum amplitude-damping channel. Phys. Rev. A \textbf{67}, 064301 (2003)

\bibitem{arshed13}Arshed, N., Toor, A.H.: Entanglement-assisted capacities of time-correlated amplitude-damping channel. arXiv:1307.5403 (2013)


\bibitem{wootters98}Wootters, W.K.: Entanglement of formation of an arbitrary state of two qubits
. Phys. Rev. Lett. \textbf{80}, 2245-2248 (1998)

\bibitem{bellomo07}Bellomo, B., Franco, R.L., Compagno, G. Non-Markovian effects on the dynamics of entanglement. Phys. Rev. Lett. \textbf{99}, 160502 (2007)






\end{thebibliography}


\end{document}